# Nanoscale capacitance: a classical charge-dipole approximation

Jun-Qiang Lu[1,2*], Jonathan Gonzalez[1], Carlos Sierra[1], Yang Li[3]

[1]Department of Physics, University of Puerto Rico, Mayaguez, PR 00681, USA

[2]Institute for Functional Nanomaterials, University of Puerto Rico, San Juan, PR 00931, USA

[3]Department of General Engineering, University of Puerto Rico, Mayaguez, PR 00681, USA

*Email: junqiang.lu@upr.edu

**ABSTRACT**

Modeling nanoscale capacitance presents particular challenge because of dynamic contribution from electrodes, which can usually be neglected in modeling macroscopic capacitance and nanoscale conductance. We present a model to calculate capacitances of nano-gap configurations and define effective capacitances of nanoscale structures. The model is implemented by using a classical atomic charge-dipole approximation and applied to calculate capacitance of a carbon nanotube nano-gap and effective capacitance of a buckyball inside the nano-gap. Our results show that capacitance of the carbon nanotube nano-gap increases with length of electrodes which demonstrates *the important roles played by the electrodes in dynamic properties of nanoscale circuits.*

## I. Introduction

Measuring electronic properties of single molecules requires nanometer spaced metallic electrodes, which are usually called a nano-gap. A nano-gap configuration, as shown in Fig. 1, is not only used widely in experimental measurements of electronic properties of single molecules[1], but also used as the standard model in theoretical modeling of quantum electron transport[2]. Recently, the nano-gap configuration was proposed to be used in DNA sequencing by measuring the difference in dc conductance of the nano-gap when different DNA nucleotides (adenine (A), cytosine (C), guanine (G), and thymine (T)) go through it[3]. However, since electronic conductance in this configuration is mainly contributed from electron tunneling from one electrode to the nucleotide and then to the other electrode, it depends exponentially on effective spaces between the nucleotide and electrodes[4]. Thus a slight difference in the effective spaces will cause huge difference in dc conductance, which makes the difference of nucleotides undistinguishable. It was reported that thousands of measurements and a good statistics might be needed in order to wash out effects of different spaces to distinguish different nucleotides[5].

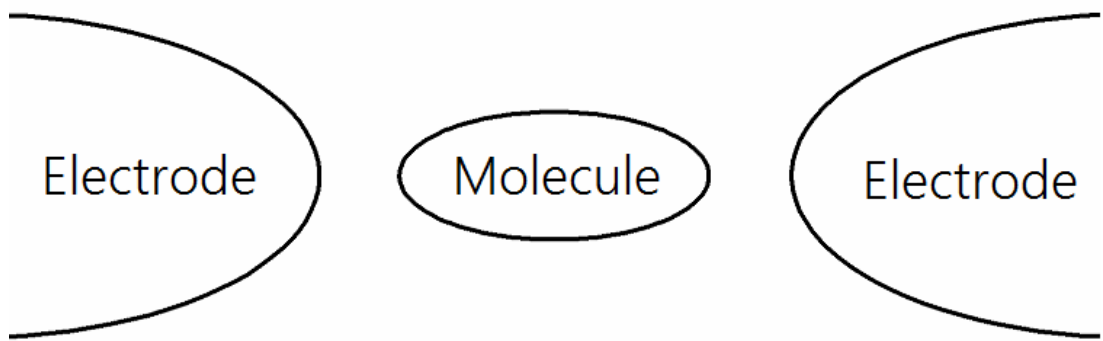

Fig. 1 A typical nano-gap configuration of two electrodes with nanometer spacing. A molecule is inside the nano-gap.

In order to alleviate the difficulties of dc conductance measurement technique, we proposed to add ac capacitance measurement[6]. Unlike conductance which is contributed from tunneling, capacitance is mainly contributed from Coulomb interaction between the nucleotide and electrodes hence is not exponentially sensitive to the effective spaces between them. Averaging over repeat measurements of capacitances may provide a better signal/noise ratio than similar averages in conductance measurements. Thus adding capacitance measurement may dramatically reduce the number of repeats.

Various theoretical models have been developed to calculate dynamic conductance[7] and capacitances[8] of nanoscale structures by extending algorithm of dc conductance, which focused on the middle device and made some assumptions to simplify the electrodes. However, in this nano-gap configuration, ac transport is very different from dc transport thus a different model needs to be developed, as illustrated in Fig. 2. In dc transport, current is *steady* not only in time, but also in position along the transport direction, as shown in Fig. 2(a). Currents through any cross-areas are the same no matter in the infinite electrodes or in the middle device. The current measured in electrodes has to go through the middle device. This is why we can use the I-V character measured in the two electrodes to study the property of the middle device. However, in ac transport, current is changing not only in time, but also in position along the transport direction, as shown in Fig. 2(b). ac currents through different cross-areas in electrodes in general are different because of charge accumulation. They will also be different from that in the middle device. In macroscopic, usually the difference can be neglected and a unified current along the electrodes can be defined due to the pretty small surface/bulk ratio. The accumulated charges are negligible when comparing with the transported charges. The current in the device can also be defined the same as the unified one by introducing the concept of displacement current. At nanoscale, with large surface/bulk ratio, the difference cannot be neglected and defining a unified current is almost impossible or otherwise questionable. Since an ac current can move back and forth, it can exist in the electrodes alone and does not have to go through the middle device. Thus the current in the electrodes can be in general much greater than that through the middle device. *Measuring dynamic I-V character in the two electrodes of a nano-gap configuration cannot tell directly the ac property of the nanoscale device in-between*. Moreover, *the* *contribution from the electrodes can be much greater than that from the middle device in modeling dynamic transport properties of nanoscale circuits*, thus we need keep in mind that *the electrodes may be more important than the middle device and need to be considered carefully*. In case one needs to find out the property of the middle device only, the contribution from the electrodes should be properly removed.

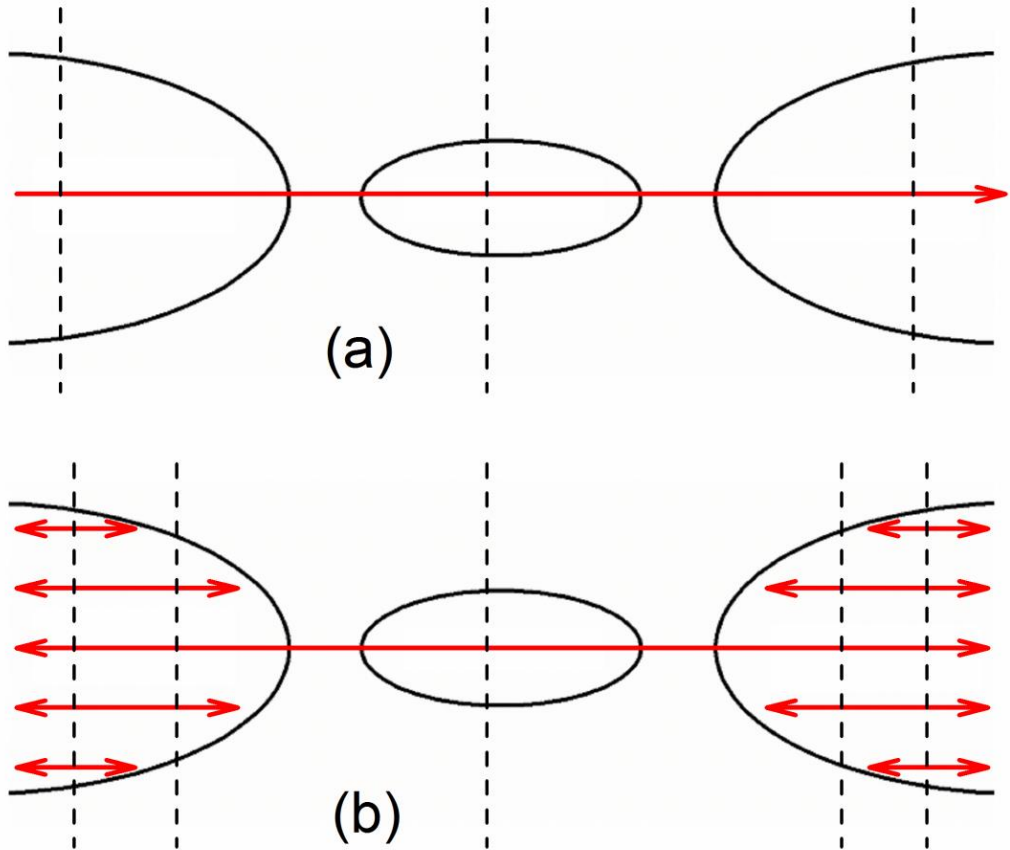

Fig. 2 (Color online) Difference between (a) dc transport and (b) ac transport at the nanoscale. The red arrows illustrate the dc and ac currents.

In this paper, we present a model to calculate capacitances of nano-gap configurations and define effective capacitances of nanoscale structures in the nano-gaps. We implement our model by using a classical atomic charge-dipole approximation[9,10] and apply it to calculate capacitance of a carbon nanotube(CNT) nano-gap and effective capacitance of a buckyball ($C_{60}$) inside the nano-gap. Our results show that the capacitance of the electrodes can be much larger than that of the middle device thus will contribute more to dynamic currents when connecting in a nanoscale circuits.

## II. Capacitances of nano-gap configurations and effective capacitances of nanoscale structures

We use the schematic drawing in Fig. 3 to represent a nano-gap configuration. In Fig. 3(a), a device is inserted into the gap between the electrodes. In Fig. 3(b) there is no device. Please note each electrode in Fig.3 has a *finite* length[11], which is just a small part of a semi-infinite electrode. In order to approach semi-infinite electrode, the electrode length will be gradually increased later in the calculations.

Since the two electrodes in Fig. 3(a) are finite in size, we can assume a finite positive charge +$Q$ accumulated on the left electrode and a finite negative charge -$Q$ accumulated on the right electrode. The potential difference $\Delta V$ between the two electrodes can then be calculated and hence the capacitance of the nano-gap can be defined as $C = Q/\Delta V$.

In order to find out the contribution from the middle device only, we can similarly calculate capacitance of the nano-gap in Fig. 3(b), $C' = Q/\Delta V'$. This is the mutual capacitance between the two electrodes only. We then define the difference between $C$ and *C'* as the effective capacitance of the device, $C_d = C - C'$. We will show below that this definition leads to a convergent $C_d$ as a function of increasing length of the electrodes.

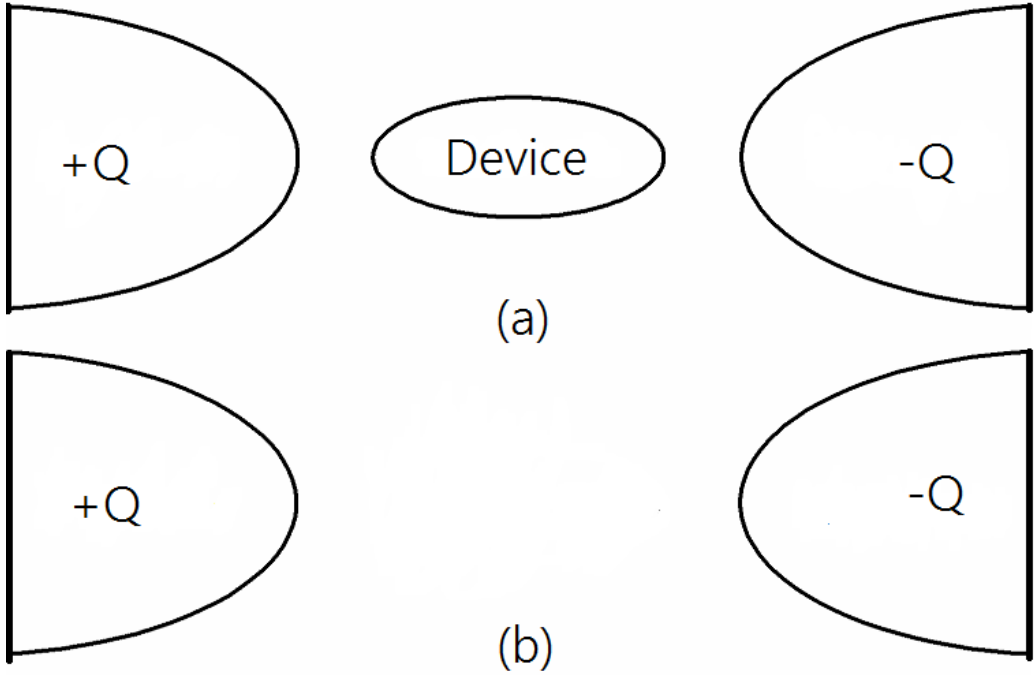

Fig. 3 A nano-gap configuration of finite-sized electrodes, (a) with and (b) without a device in-between.

### III. Classical Charge-Dipole approximation

The model in Section II requires calculations of the potential difference $\Delta V$ between the two electrodes when one has +*Q* and the other has -*Q*. To find out the potential difference, we use an atomic Charge-Dipole approximation, which has been successfully used in carbon nanostructures[9,10]. The charge distribution of each atom is approximated by a charge *q* and a dipole **p**, which are assumed to be Gaussian distributions. Thus the charge density $\rho_j(\mathbf{r})$ of atom *j* includes two parts: $\rho(\mathbf{r})[q_j] = (q_j/\pi^{3/2} R^3)e^{-(|\mathbf{r}-\mathbf{r}_j|/R)^2}$ and $\rho(\mathbf{r})[\mathbf{p}_j] = \mathbf{p}_j \cdot \nabla \rho(\mathbf{r})[1]$, where **r**$_j$ is the position of the atom *j* and *R* is the width of the distributions, $\rho(\mathbf{r})[1] = (1/\pi^{3/2}R^3)e^{-(|\mathbf{r}-\mathbf{r}_j|/R)^2}$. The charge $q_j$ and dipole $\mathbf{p}_j$ can be determined by solving the following equations:

$$\begin{cases} \sum_{j=1}^{N} \mathbf{T}_{ij}^{pp} \otimes \mathbf{p}_j + \sum_{j=1}^{N} \mathbf{T}_{ij}^{pq} q_j = 0; \\ \sum_{j=1}^{N} \mathbf{T}_{ij}^{pq} \cdot \mathbf{p}_j + \sum_{j=1}^{N} T_{ij}^{qq} q_j + \chi_i - V_L = 0; i \in L \\ \sum_{j=1}^{N} \mathbf{T}_{ij}^{pq} \cdot \mathbf{p}_j + \sum_{j=1}^{N} T_{ij}^{qq} q_j + \chi_i - V_R = 0; i \in R \\ \sum_{j \in L} q_j = +Q; \\ \sum_{j \in R} q_j = -Q, \end{cases}$$

where $T_{ij}^{qq} = (1/4\pi\epsilon_0)\mathrm{erf}(r_{ij}/\sqrt{2}R)/r_{ij}$, $\mathbf{T}_{ij}^{pq} = -\nabla_{r_i} T_{ij}^{qq}$, $\mathbf{T}_{ij}^{pp} = \nabla_{r_j} \otimes \mathbf{T}_{ij}^{pq}$, N is the total number of atoms, $\chi_i$ is the electron affinity of atom $i$, $i \in L(R)$ means atom $i$ belongs to left(right) electrode, $V_L$ and $V_R$ are electric potential of the left and right electrodes which can also be determined by solving the above equations. After determining $V_L$ and $V_R$, the potential difference $\Delta V = V_L - V_R$.

## IV. Capacitance of a CNT nano-gap and effective capacitance of C$_{60}$ in it

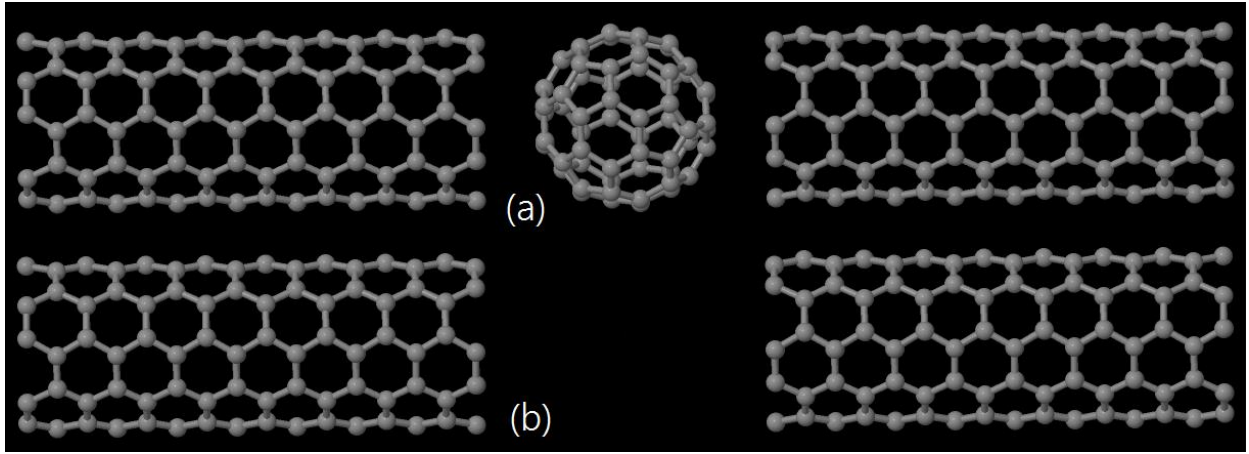

Fig. 4 A nano-gap of CNT(5,5), (a) with and (b) without a $C_{60}$ in between.

The above model and method are then applied to study the capacitance of a CNT nano-gap and effective capacitance of a $C_{60}$ in the nano-gap. As shown in Fig. 4(a), we are using a nano-gap which includes two (5,5) CNTs. The gap between them is 1.23 nm. To apply the method, we use electron affinity of carbon atom $\chi = 1.26$ eV, and width of Gaussian distributions $R$ = 0.06862nm[9]. We then assume the charge on the left electrode is $+e$ and that on the right electrode is $-e$, and calculate the charge distributions and determine potential difference between the two electrodes. Fig. 5 shows the calculated charge distributions when each electrode has a length of 8 unit cells ($N_C$ = 8). Clearly, the $C_{60}$ in the gap is positively charged on one side and negatively charged on the other side due to the charges in the electrodes. The charges on the $C_{60}$ then will change the charge distributions and potential profiles in the electrodes through Coulomb interaction and hence contribute to the capacitance of the nano-gap. This contribution is the earlier defined effective capacitance of middle device. After determining potential difference between the two electrodes, the capacitance of the nano-gap is then calculated by $C = e/\Delta V$. Fig. 6(a) presents the calculated capacitance as a function of the length of electrodes $N_C$. By increasing the length of electrodes, the capacitance keeps going up, *which shows the importance to properly include electrodes in studying dynamic transport properties.* In order to find out the contribution from the $C_{60}$, we also calculate the $C'$ which is the capacitance of the CNT nano-gap without the $C_{60}$ in-between as shown in Fig. 4(b). The effective capacitance $C_d$ of the $C_{60}$ is then calculated as the difference between $C$ and $C'$ and plotted in Fig. 6(b). Obviously the effective capacitance $C_d$ also increases with the length of electrodes at the beginning, however, it converges very quickly. Longer than 64 unit cells, the effective capacitance of $C_{60}$ converges to 0.03532 e/V, which is much smaller than the capacitance of the electrodes. Please notice that the converged value of $C_{60}$ is still electrode-dependent. It will change when changing to different electrodes. It is the dielectric effect of $C_{60}$ on the capacitance of the nano-gap. That's why we call it an effective capacitance of the device.

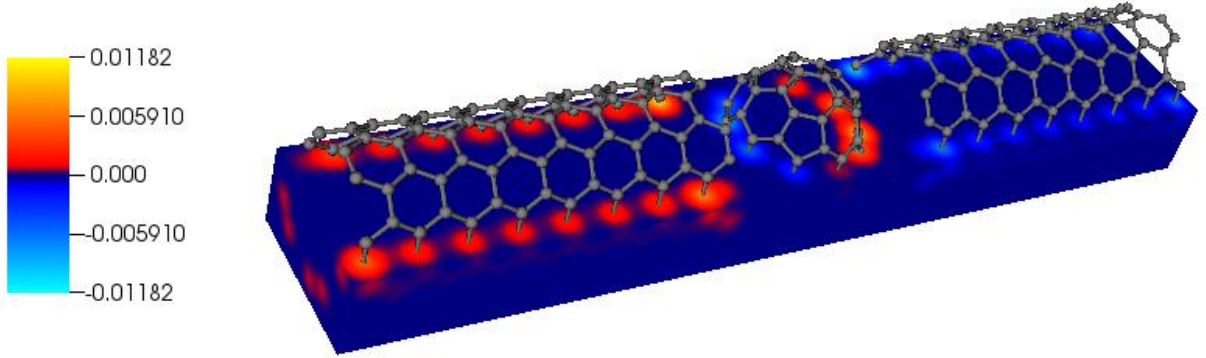

Fig. 5 (Color online) Charge distributions of the CNT nano-gap and in-between $C_{60}$ with $+e$ on one electrode and and $-e$ on the other.

## V. Conclusions

We present a model to calculate capacitances of nano-gap configurations and define effective capacitances of nanoscale structures. By assuming charge accumulation in the two electrodes of a nano-gap, the capacitance of the nano-gap is calculated by determining potential difference between the two electrodes. The effective capacitance of a nanoscale structure is then defined by the difference between the two capacitances of the nano-gap with and without the structure in between. We implement the model by using a classical atomic charge-dipole approximation and apply it to calculate capacitance of a CNT nano-gap and effective capacitance of a $C_{60}$ inside the nano-gap. Our results show that the capacitance of the CNT nano-gap increases with the length of electrodes and the effective capacitance of the $C_{60}$ reaches a converged value at certain length of electrodes. Moreover, the converged effective capacitance of the $C_{60}$ is much smaller than the capacitance of the CNT nano-gap, which demonstrates *the importance to consider the contribution of electrodes in studying dynamic transport properties of nanoscale circuits.*

*This research is supported by an award from Research Corporation for Science Advancement.*

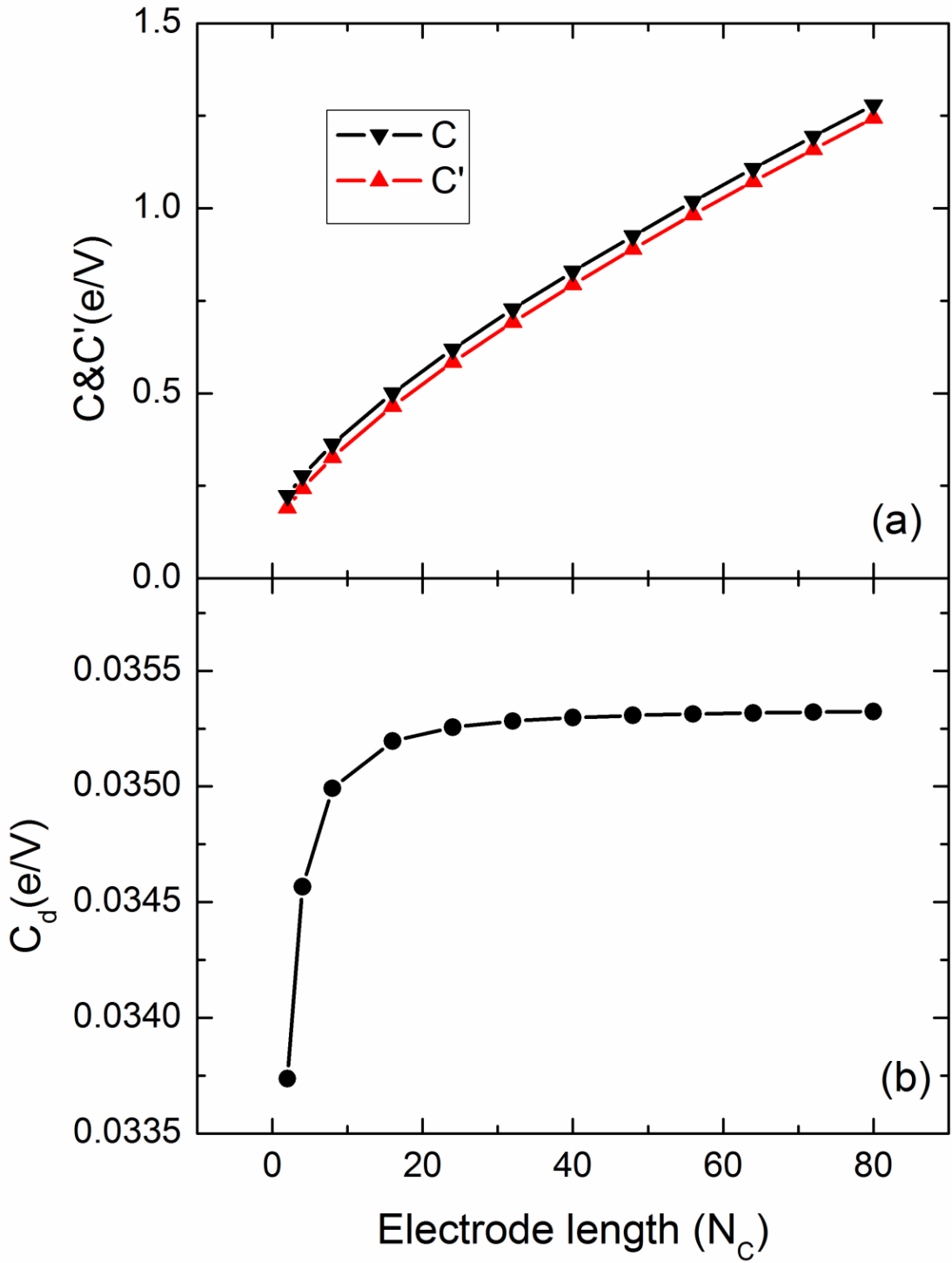

Fig. 6 (Color online) (a) Capacitance of a CNT nano-gap with a $C_{60}$ in-between or not, and (b) effective capacitance of the $C_{60}$, as functions of electrode length.